\documentclass[12pt]{article}

\pdfoutput=1

\usepackage{slashed}
\usepackage{amsmath,amssymb,graphicx,multicol} 
\usepackage{epsf,color}
\usepackage{pstricks}
\usepackage{cite}

\newcommand{\beq}{\begin{eqnarray}}
\newcommand{\eeq}{\end{eqnarray}}

\newcommand{\centeron}[2]{{\setbox0=\hbox{#1}\setbox1=\hbox{#2}\ifdim

\wd1>\wd0\kern.5\wd1\kern-.5\wd0\fi \copy0

\kern-.5\wd0\kern-.5\wd1\copy1\ifdim\wd0>\wd1
                                       \kern.5\wd0\kern-.5\wd1\fi}}
\newcommand{\ltap}{\>\centeron{\raise.35ex\hbox{$<$}}
                               {\lower.65ex\hbox{$\sim$}}\>}
\newcommand{\gtap}{\>\centeron{\raise.35ex\hbox{$>$}}
                               {\lower.65ex\hbox{$\sim$}}\>}

\newcommand\ZZ{\hbox{\zfont Z\kern-.4emZ}}
\font\zfont = cmss10 

\renewcommand{\theequation}{\thesection.\arabic{equation}}
\begin{document}
\begin{titlepage}
\begin{flushright}
\end{flushright}

\vskip.5cm

\begin{center}
{\huge \bf Anomalies, Unparticles, } 
\vskip.2cm
{\huge \bf and Seiberg Duality}
\end{center}
\vskip.1cm
\begin{center}

\vskip.1cm
\end{center}
\vskip0.2cm

\begin{center}
{\bf Jamison Galloway, John McRaven and John Terning} 
\end{center}
\vskip 8pt

\begin{center}
{\it Department of Physics, University of California, Davis,
CA
95616.} \\
\vspace*{0.3cm}
{\tt  galloway@physics.ucdavis.edu, mcraven@physics.ucdavis.edu, terning@physics.ucdavis.edu}
\end{center}

\vglue 0.3truecm

\begin{abstract}
\vskip 3pt \noindent
We calculate triangle anomalies for fermions with non-canonical scaling dimensions.  
The most well known example of such fermions  (aka unfermions) occurs in Seiberg duality where the matching of anomalies (including mesinos with scaling dimensions between 3/2 and 5/2) is a crucial test of duality. By weakly gauging the non-local action for an unfermion, we  calculate the one-loop three-current amplitude.  Despite the fact that there are  more graphs with more complicated propagators and vertices, we find that the calculation can be completed in a way that nearly parallels the usual case.  We show that the anomaly factor for fermionic unparticles is independent of the scaling dimension  and identical to that for ordinary  fermions. This can be viewed as a confirmation that unparticle actions correctly capture the physics of conformal fixed point theories like Banks-Zaks or SUSY QCD.
\end{abstract}

\end{titlepage}

\renewcommand{\theequation}{\thesection.\arabic{equation}}
\newcommand{\ns}{\negthinspace}
\section{Introduction}\label{sec:Introduction}
\setcounter{equation}{0}
\setcounter{footnote}{0}
Recently Georgi introduced unparticles as a simple way of dealing with fields that have non-canonical scaling dimensions that are part of a conformal theory \cite{Georgi1, Georgi2}. The  phase space for these  unparticles formally resembles that of a non-integer number of particles, hence the unusual name.
Subsequent studies have examined conformal symmetry breaking \cite{shirmanquiros}, how gauge interactions can be introduced  \cite{uncolor}, and how such theories are related to the AdS/CFT correspondence, in particular how their effective actions relate to holgraphic boundary actions \cite{unfermions}.

Since unparticles have unusual propagators (as dictated by conformal symmetry) and gauge interactions, one obvious question is  how to reproduce some of the basic aspects of quantum field theories such as the triangle anomaly. It is well known that a theory containing both vector (V) and axial-vector (A) symmetries cannot in general preserve both upon quantization.  This was first demonstrated in perturbation theory by Adler, Bell, and Jackiw \cite{ABJ}.  In the language of path integrals, the anomaly  arises from a symmetry that is respected by the action but not by the integration measure.  Fujikawa has shown \cite{Fujikawa} that the effect of axial transformations on the fermionic path integral measure leads to the same anomaly factor as that derived diagrammatically.  The absence of gauge anomalies is  necessary for the consistency of a gauge theory, thus in order to talk about gauge interactions for unparticles it is necessary to be able to show that the gauge anomaly vanishes.

Anomalies are also an ensential test of  Seiberg's dual description of SUSY QCD \cite{Seiberg}. Seiberg used 't Hooft anomaly matching \cite{tHooft} to check whether the two dual descriptions described the same physics.  When the global symmetries of SUSY QCD are unbroken, the anomalies exactly match onto a dual description which includes a chiral superfield  ``meson".  For an $SU(N)$ gauge group in the conformal window, where the number of quark/squark flavors is between $3N/2$  and $3 N$, the anomaly matching relates the anomaly factors of fermions with non-canonical scaling dimensions, i.e. in Georgi's language the anomalies of unfermions.  This should be a valid proceedure since the anomaly is not renormalized at any finite order of perturbation theory \cite{AdlerBardeen} and we
expect that the anomaly should be independent of scaling dimension.  In this paper we will show how this comes about directly from the exact fermion propagators that are required by conformal invariance.

 In Sec.~2 we review how to gauge the unfermion action and the derivation of the Feynman rules for unfermions.  In Sec.~3 we use these Feynman rules to explicitly calculate the anomaly  diagrams involving unfermions and show that the result matches the canonical result exactly: the anomaly factors contain no dependence on the scaling dimension of the unfermion.  
We conclude  with some speculations in Sec.~4.

\section{Gauge Interactions and Unfermions}\label{sec:Rules}
\setcounter{equation}{0}
\setcounter{footnote}{0}
The potentially large anomalous dimension of an unparticle operator may be viewed as the effect of gauge bosons (or other massless particles) from a CFT which are implicitly summed in the unparticle two-point function. For this reason, propagating unparticles bear a close resemblance to QCD jets, as shown in \cite{Neubert}.  So unparticles seem to be a natural way to try to understand interacting conformal fixed points in gauge theories, especially when we need to see the effects of weakly gauging a global symmetry of the (possibly strongly interacting) gauge theory. The question of whether it is safe to neglect possible mixings between external gauge bosons and spin one composites of the CFT  has been addressed in Ref. \cite{unfermions}, but  is not strictly relevant here since we are primarily interested in the limit of zero external gauge coupling which gives the  anomaly for a global symmetry.  

Recall that the unitarity bound \cite{Mack,reviews} on the scaling dimension of a gauge invariant spin 1/2 operator is
\beq
d\ge \frac{3}{2}~,
\label{unitarity}
\eeq
with equality holding for a free field.
 In the conformal window, SUSY QCD and Seiberg's dual description are superconformal so the scaling dimension of  chiral superfields is simply related to the superconformal $R$-charge \cite{Flato,reviews} via
\beq
d[{\mathcal O}]=\frac{3}{2}|R_{sc}[{\mathcal O}]|.
\label{rcharge}
\eeq
For $N$ colors and $F$ flavors the superconformal $R$-charge fixes the dimension of the scalar meson operator to be
\beq
d_s=3-\frac{3N}{F}~,
\eeq
while the its fermionic superpartner, the mesino, has scaling dimension
\beq
d=\frac{7}{2}-\frac{3N}{F}~.
\label{mesinodim}
\eeq
We see from (\ref{mesinodim}) that for $F=3N$, $d=5/2$ and that $d$ decreases with $F$.  As the number of flavors approaches $3N/2$, $d$ approaches the unitarity bound (\ref{unitarity}), and the mesino becomes a free field at $F=3N/2$.
Thus Seiberg's theory provides us with a simple, well-known example of a theory containing unfermions with dimensions between $3/2$ and $5/2$.

Now let us review the procedure for constructing a gauge theory with unfermions as in \cite{unfermions} (see also \cite{Basu}).  Using the generalized form for a spectral decomposition, the unfermion  propagator is determined using dimensional analysis as it was  for unbosons with various spins \cite{Georgi1, Georgi2,Cheung,Grinstein}.   Taking the unfermions to have scaling dimension $d$, the following expression for the propagator with  $3/2 \leq d<5/2$ is required by conformal symmetry (up to a normalization):
\beq\label{prop ansatz}
\Delta(p)&=&\frac{A_f}{2 \pi}\int_{0}^{\infty}dM^2(M^2)^{d-5/2}\frac{i \,{\slashed p}}{p^2-M^2+i \epsilon} \\
&=& \frac{A_f}{2 i \cos (d \pi)}\,{\slashed p}\, (-p^2-i \epsilon)^{d-5/2}. \nonumber
\eeq
Note we are using Dirac notation for the fermion although we really have independent left- and right-handed (Weyl) unfermions.  We will use Dirac spinors  in order  to simplify comparisons to the traditional calculation. 

The normalization is taken to be
\beq
A_f(d)=\frac{16 \pi^{5/2}}{(2 \pi)^{2d-1}}\frac{\Gamma(d)}{\Gamma(d-3/2)\Gamma(2d-1)}
\eeq
so that the regular propagator for a fermion is reproduced when we take the limit of a canonical (free) scaling dimension: $d\rightarrow 3/2$.  

To derive the Feynman rules for the unfermions, we first write the effective momentum-space action which gives rise to Eq.~(\ref{prop ansatz}):
\beq\label{p action}
S&=& \frac{-2 \cos(d \pi)}{A_f}\int \frac{d^4 p}{(2 \pi)^4} \ {\bar \psi}  (p) \frac{(-p^2)^{5/2-d}}{{\slashed p}} \psi (p)  \\
&=&\frac{2 \cos (d \pi)}{A_f} \int \frac{d^4 p}{(2 \pi)^4} \ {\bar \psi}  (p) (-p^2)^{3/2-d}\,{\slashed p}\, \psi (p) \nonumber \\
&\equiv&\frac{2 \cos (d \pi)}{A_f} \int \frac{d^4 p}{(2 \pi)^4} \ {\bar \psi}  (p) F(p)\ {\slashed p}\ \psi (p)\nonumber
\eeq
where $F(p)$ has been introduced in order to facilitate comparisons with well-known  results.  The unfermion propagator in this notation (neglecting the $+i \epsilon$ term) is:
\beq
\Delta(p)=\frac{A_f}{2 \cos(d \pi)}\frac{i}{{\slashed p}F(p)}.
\eeq

For the calculation of the anomaly, we will consider weakly gauging vector and axial Abelian symmetries, although ultimately only the vector symmetry can be gauged.    To accommodate the gauge symmetry we Fourier transform Eq.~(\ref{p action}) into a non-local position space action and introduce a Wilson line to ensure that this action is independent of gauge transformations.  This gives the following gauge-invariant action (we momentarily retain non-Abelian group generators for generality in these expressions):
\beq\label{gauged action}
S=\frac{2\cos(d \pi)}{A_f} \int d^4x \ d^4y \ {\bar \psi}(x) \left(-i \slashed \partial_y {\tilde F}(x-y)\right) W(x,y) \psi(y),
\eeq
where
\beq\label{Wilson lines}
W(x,y)&=&\exp \left[-ig T^a  \int_x^y dw^\mu V_\mu^a(w) -ig T^a \gamma^5 \int_x^y dw^\mu A_\mu^a(w)\right], 
\eeq
and the derivative acts on ${\tilde F}(x-y)$.

We can now determine the vertex functions\footnote{The method has  previously been applied in the case of non-local toy models \cite{NL, NCQM}.} by taking functional derivatives of Eq.~(\ref{gauged action}) with respect to appropriate combinations of fields.  This is exactly as in \cite{unfermions} and can be shown to be a natural extension of the usual minimal coupling prescription (cf. \cite{minimal}).   The vertex function for two unfermions and one gauge boson is then given by:
\beq
ig\Gamma^{\mu a}(p,q)&=&\frac{i\delta^3 S}{\delta V_{\mu}^a(q)\delta {\bar \psi}(p+q)\delta  \psi  (p)}  \\ 
&=&i g  \left\{\frac{1}{2}\gamma^{\mu}T^a\left[F(p+q)+F(p)\right]\right. \nonumber \\ && \qquad  +\left.\left(\slashed p + \frac{1}{2}\slashed q\right)T^a\frac{(2p+q)^{\mu}}{2p\cdot q +q^2} \left[F(p+q)-F(p)\right] \right\}. \nonumber
\eeq
(Here and below we suppress normalization factors as they will cancel in the calculation of observable quantities).  In our calculation with Abelian symmetries we can replace all group generators by the identity. For the vertex involving an axial gauge boson we find 
\beq\label{axial vertex1}
ig \Gamma^{\mu_5} = ig \Gamma^\mu \gamma^5.
\eeq  
One can check that these vertex functions satisfy the Ward-Takahashi (W-T) identity \cite{WT}:
\beq\label{WT1}
iq_\mu \Gamma^{\mu a}(p,q)=[\Delta^{-1}(p)-\Delta^{-1}(p+q)]T^a.
\eeq
We will also need the explicit form of the vertex involving two unfermions and two gauge bosons:
\beq
&& \ns \ns \ns \ns \ns \ns ig^2\Gamma^{\mu a, \nu b}  (p,q_1,q_2) =  \frac{i\delta^3 S}{\delta V_{\mu}^a(q_1)\delta V_\nu^b (q_2) \delta {\bar \psi}(p+q_1+q_2)\delta  \psi  (p)} \\
&=& \ns \ns \frac{ig^2}{2}\Bigg\{(2{\slashed p}+{\slashed q_1}+{\slashed q_2})\Big[(T^aT^b+T^bT^a)g^{\mu \nu}{\mathcal F}(p,p+q_1+q_2)\nonumber \\
&&\quad\quad+ \ns T^bT^a\frac{(2 p^{\mu}+q_1^{\mu})(2p^\nu+2q_1^\nu+q_2^\nu)}{(p+q_1+q_2)^2-(p+q_1)^2}\left({\mathcal F}(p,p+q_1+q_2)-{\mathcal F}(p,p+q_1)\right)\nonumber \\
&&\quad\quad+ \ns T^aT^b\frac{(2 p^{\nu}+q_2^{\nu})(2p^\mu+q_1^\mu+2q_2^\mu)}{(p+q_1+q_2)^2-(p+q_2)^2}\left({\mathcal F}(p,p+q_1+q_2)-{\mathcal F}(p,p+q_2)\right)\Big]\nonumber \\
&&\quad\quad+ \ns \gamma^\mu T^aT^b(2p^\nu+q_2^\nu){\mathcal F}(p,p+q_2)+\gamma^\mu T^bT^a (2p^\nu+2q_1^\nu+q_2^\nu){\mathcal F}(p+q_1,p+q_1+q_2)\nonumber \\
&&\quad\quad+ \ns \gamma^\nu T^bT^a(2p^\mu+q_1^\mu){\mathcal F}(p,p+q_1)+\gamma^\nu T^aT^b(2p^\mu+q_1^\mu+2q_2^\mu){\mathcal F}(p+q_2,p+q_1+q_2)\Bigg\} \nonumber
\eeq
where the definition
\beq
\mathcal{F}(p_1,p_2)\equiv \frac{F(p_1)-F(p_2)}{p_1^2-p_2^2}
\eeq
elucidates the canonical $d=3/2$ limit where $F(p)\rightarrow1$ and  this vertex vanishes.  We can check that this vertex also satisfies a W-T identity given by
\beq
i  q_{1\mu} \Gamma^{a\mu, b\nu} (p,q_1,q_2) &=&  i  \Gamma^{b\nu} (p+q_1, q_2) T^a - i  T^a \Gamma^{b\nu} (p, q_2)   \nonumber \\
& & - f^{abc}\,   \Gamma^{c\nu} (p, q_1+q_2).
\eeq
which in the Abelian case simplifies to 
\beq
i q_{1 \mu}\Gamma^{\mu \nu}(p,q_1,q_2)=i\Gamma^{\nu}(p+q_1,q_2)-i\Gamma^{\nu}(p,q_2).
\eeq
Finally, the VA vertex function is given by 
\beq
\Gamma^{\mu_5,\nu}=\Gamma^{\mu \nu}\gamma^5.
\eeq

In the unfermion case it might seem possible to generate a VVA loop amplitude with a single VVA vertex attached to a fermion loop (as shown in the last diagram of figure 1).  However  the fermion loop requires a trace which will yield a non-zero result only with an appropriate number of gamma matrices together with the $\gamma^5$.  The diagram with three gauge bosons at a single vertex has a factor containing a single $\gamma^\mu \gamma^5$ from the vertex, multiplied by one additional gamma matrix from the diagram's single propagator.  Thus this  diagram  vanishes.

\section{Calculating the Anomaly}
\setcounter{equation}{0}

In the unfermion case there are altogether six VVA diagrams, all shown in Figure 1.
\begin{figure}[ht]
\centering
\includegraphics[width=6cm]{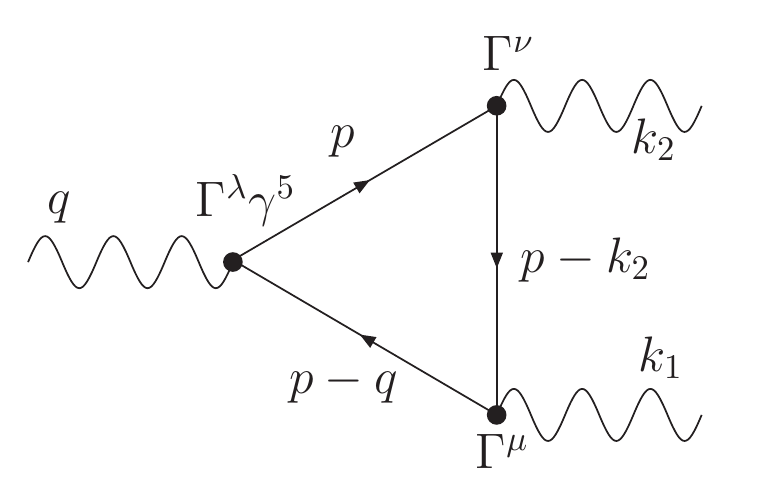}\hspace{40pt} \includegraphics[width=6.5cm]{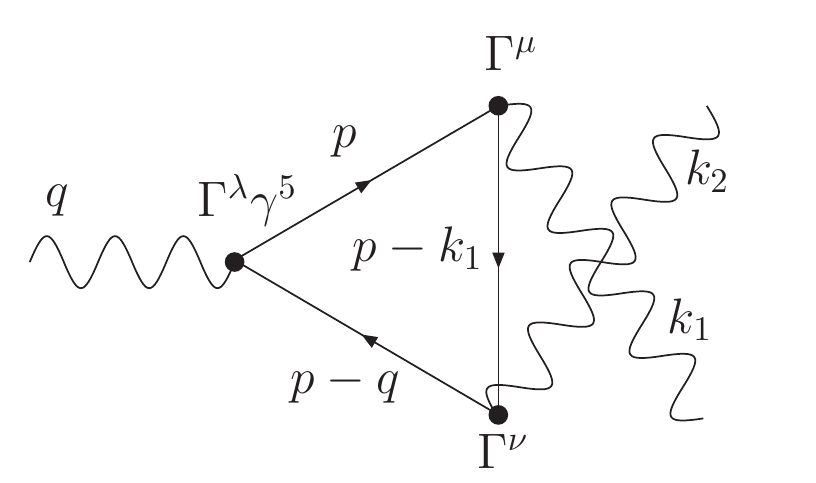} \\
\includegraphics[width=6.5cm]{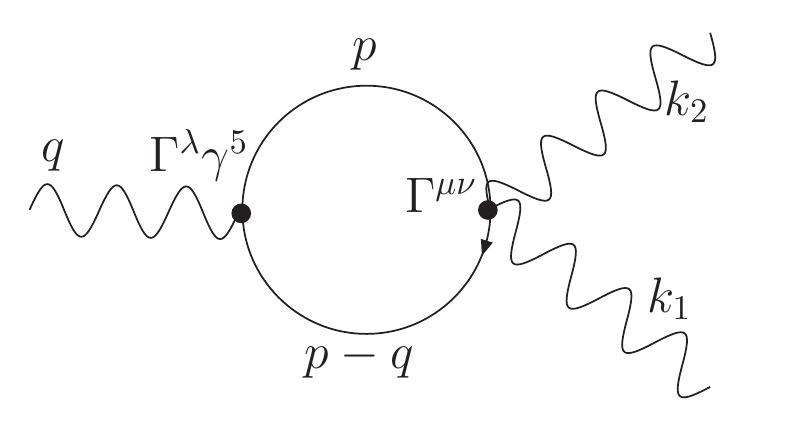} \hspace{40pt} \includegraphics[width=6cm]{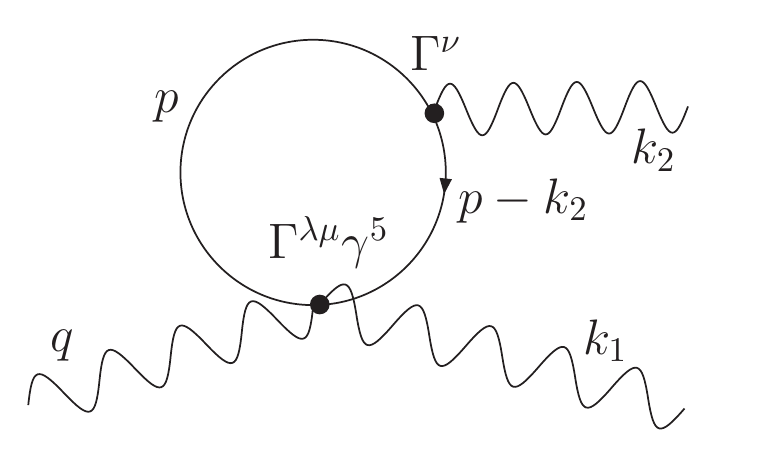} \\
\includegraphics[width=6cm]{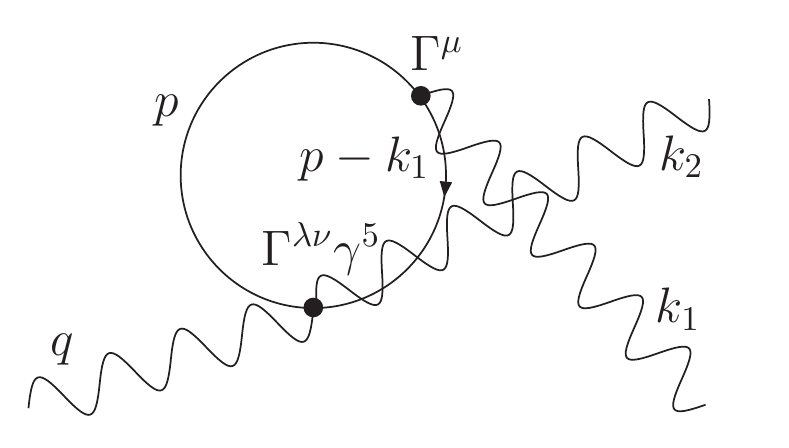} \hspace{40pt} \includegraphics[width=6cm]{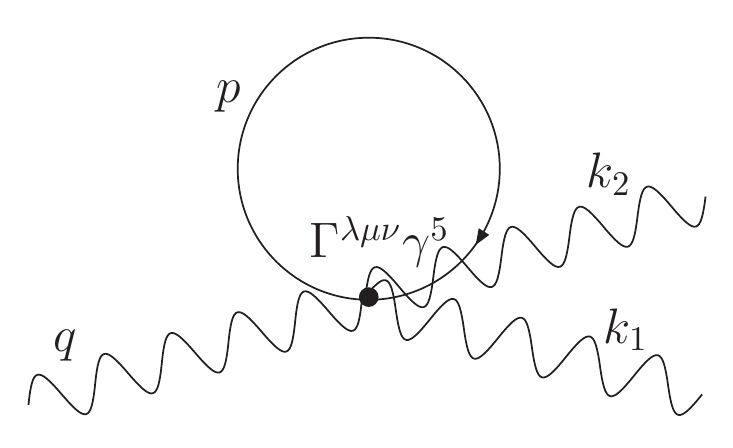} \caption{VVA diagrams for unfermions}
\end{figure}
From these we calculate the total amplitude $T^{\lambda \mu \nu}(q,k_1,k_2;p)$; the loop momentum routing is made explicit as  the final result  depends upon it.   As shown in the diagrams we take momentum $q$ to correspond to  the axial current, while $k_1$ and $k_2$ correspond to the vector currents.  The full expression for the amplitude is given by:
\beq\label{amplitude}
T^{\lambda \mu \nu}(q,k_1,k_2;p)&=& \nonumber \\ 
-\int \frac{d^4p}{(2 \pi)^4} \ {\rm tr} \hspace{-10pt}  &\Big[& \hspace{-12pt} \Gamma^{\lambda_5}(p-q,q)\Delta(p-q) \Gamma^\mu(p-k_2,-k_1)\Delta(p-k_2)\Gamma^\nu(p,-k_2)\Delta(p) \nonumber \\
&& \hspace{-12pt} +\Gamma^{\lambda_5}(p-q,q)\Delta(p-q) \Gamma^\nu(p-k_1,-k_2)\Delta(p-k_1)\Gamma^\mu(p,-k_1)\Delta(p) \nonumber \\
&& \hspace{-12pt} +\Gamma^{\lambda_5}(p-q,q)\Delta(p-q)\Gamma^{\mu \nu}(p,-k_1,-k_2)\Delta(p) \nonumber \\
&& \hspace{-12pt} +\Gamma^{\lambda_5 \mu}(p-k_2,q,-k_1)\Delta(p-k_2)\Gamma^\nu(p,-k_2)\Delta(p) \nonumber \\
&& \hspace{-12pt} +\Gamma^{\lambda_5 \nu}(p-k_1,q,-k_2)\Delta(p-k_1)\Gamma^\mu(p,-k_1)\Delta(p) \nonumber \\
&& \hspace{-12pt} +\Gamma^{\lambda_5\mu \nu}(p,q,-k_1,-k_2)\Delta(p)\Big].
\eeq

The Fourier transform of the divergence of the axial currents is thus  $q_\lambda T^{\lambda \mu \nu}(q,k_1,k_2)$.  Note that not all terms in Eq.~(\ref{amplitude}) will contribute to the trace; we find, after using the appropriate W-T identities, that the only terms which will give non-zero contributions are those containing two propagators and two single gauge boson vertices, since the Dirac trace with a $\gamma^5$ vanishes  unless there are four additional $\gamma^\mu$'s.  Using $q=k_1+k_2$ we have
\beq
q_\lambda T^{\lambda \mu \nu}(q,k_1,k_2;p) &=&\nonumber \\
g  \int \frac{d^4 p}{(2 \pi)^4} \ {\rm tr} \Big[\hspace{-10pt}&&\hspace{-18pt}\gamma^5 \Gamma^\mu(p-k_2,-k_1)\Delta(p-k_2)\Gamma^\nu(p,-k_2)\Delta(p)  \nonumber \\
&&\hspace{-20pt}+ \gamma^5 \Delta(p-q) \Gamma^\mu(p-k_2,-k_1)\Delta(p-k_2)\Gamma^\nu(p,-k_2) \nonumber \\
&+&\hspace{-10pt}\gamma^5 \Gamma^\nu(p-k_1,-k_2)\Delta(p-k_1)\Gamma^\mu(p,-k_1)\Delta(p) \nonumber \\
&+&\hspace{-10pt} \gamma^5 \Delta(p-q) \Gamma^\nu(p-k_1,-k_2)\Delta(p-k_1)\Gamma^\mu(p,-k_1) \nonumber \\
&+&\hspace{-10pt} \gamma^5\Delta(p-k_2)\Gamma^\nu(p,-k_2)\Delta(p)\Gamma^\mu(p-k_2,-k_1) \nonumber \\
&-&\hspace{-10pt} \gamma^5\Delta(p-k_2)\Gamma^\nu(p,-k_2)\Delta(p)\Gamma^\mu(p+k_1,-k_1) \nonumber \\
&+&\hspace{-10pt} \gamma^5\Delta(p-k_1)\Gamma^\mu(p,-k_1)\Delta(p)\Gamma^\nu(p-k_1,-k_2) \nonumber \\
&-&\hspace{-10pt} \gamma^5\Delta(p-k_1)\Gamma^\mu(p,-k_1)\Delta(p)\Gamma^\nu(p+k_2,-k_2)\Big].
\eeq

From this point, the calculation closely follows the canonical  case, though the cancellation between diagrams necessarily includes more than just the diagrams present in the $d=3/2$ limit.  Indeed, using the cyclic property of the trace and the anti-commutation of $\gamma^5$ with the other gamma matrices, we see that the first and fifth terms will cancel exactly, as will the third and seventh.  We also find that the sixth term can be generated from the fourth (up to the sign difference) by taking $p\rightarrow p+k_1$; similarly the second term generates the eighth by taking $p\rightarrow p+k_2$.  Taking account of these cancellations we write the result in the following suggestive form:
\beq \label{integralform}
q_\lambda T^{\lambda \mu \nu}(q,\hspace{-13pt} && \hspace{-13pt} k_1,k_2;p) = \nonumber \\
g  \int \hspace{-13pt} && \hspace{-13pt} \frac{d^4 p}{(2 \pi)^4}  {\rm tr} \Big[\gamma^5 \Delta(p-q) \Gamma^\nu (p-k_1,-k_2)\Delta (p-k_1) \Gamma^\mu(p,-k_1)  \nonumber \\
\hspace{-30pt}&-& \gamma^5 \Delta(p-k_2) \Gamma^\nu (p,-k_2)\Delta (p) \Gamma^\mu(p+k_1,-k_1)+\{k_1 \leftrightarrow k_2; \mu \leftrightarrow \nu\}\Big] \nonumber \\
&\equiv& -g  \int \frac{d^4 p}{(2 \pi)^4} \big[f(p+k_1)-f(p) +\{k_1 \leftrightarrow k_2; \mu \leftrightarrow \nu \}\big],
\eeq
with
\beq
f(p) \equiv {\rm tr} \left[\gamma^5 \Delta(p-q) \Gamma^\nu (p-k_1,-k_2)\Delta (p-k_1) \Gamma^\mu(p,-k_1)\right].
\eeq
We can now use the usual method  of applying Gauss's theorem \cite{Bertlmann}, i.e. take
\beq
\int \frac{d^4p}{(2\pi)^4} \left[f(p+k)-f(p)\right] = \lim_{P\rightarrow \infty} \frac{i k^\mu P_\mu P^2}{8 \pi^2} f(P)
\eeq
so that we'll need to consider only those terms from $f(p)$ that are of ${\mathcal O} (p^{-3})$, that is the linearly divergent terms.  With this only  the following terms remain:
\beq 
f(p) &\supset& g^2 \left\{ \frac{i \epsilon^{\mu \nu \alpha \beta} k_{2 \alpha} p_\beta}{p^2(p-k_2)^2}  \left[(p-k_2)^{2d-3}p^{3-2d}+(p+k_1)^{3-2d}p^{2d-3}\right. \right. \nonumber \\
&& \hspace{110pt} +\left. (p-k_2)^{2d-3}(p+k_1)^{3-2d}+1\right]\Bigr\}.
\eeq
Thus in the limit $p\rightarrow \infty$, the only contribution is:
\beq
\frac{4ig^2  \epsilon^{\mu \nu \alpha \beta} k_{2 \alpha} p_\beta}{p^4}.
\eeq
Taking the average $p^\mu p^\nu/p^2\rightarrow \eta^{\mu \nu}/4$, we have (now accounting for the $k_1, \mu \leftrightarrow k_2, \nu$ terms)
\beq
q_\lambda T^{\lambda \mu \nu}(q, k_1, k_2;p) = \frac{g^3}{4 \pi^2} \epsilon^{\mu \nu \alpha \beta}k_{1\beta} k_{2 \alpha}.
\eeq
The contriubtions to the divergence of the vector currents  will all lead to similar results, in particular  that the pertinent quantities are independent of the scaling dimension of the unfermion.  Provided we can cast all linearly divergent integrals into the form \beq
\int dp \ [f(p+k)-f(p)]~,
\eeq
we will only ever have to consider limits of integrands as $p\rightarrow \infty$.   Writing the integrals in this form is however quite non-trivial as we must include the contributions from all of the nonstandard diagrams.   Moreover, this step will remove any dependence on the scaling dimension as can be seen by examining the vertex functions: in this limit, factors of $F(p+k)$, where $k$ represents any other momentum in the problem, are replaced by $F(p)$.  When this replacement is made, we find
\beq
ig \Gamma^\mu(p,q) \rightarrow ig \gamma^\mu F(p)~,
\eeq
while any vertex function with more than one gauge boson vanishes.  Thus each contribution will have one numerator power of $F(p)=(-p^2)^{3/2-d}$ coming from each vertex function and one denominator power of the same function involving $p$ coming from each propagator.  Thus as we consider the limit $p\rightarrow \infty$ we will lose any dependence on $d$ as the $F(p)$'s --- which  encode {\it all} of the $d$-dependence in the problem --- will cancel.

We can follow similar steps to those above to calculate the vector current divergence, $k_{1\mu}T^{\lambda \mu \nu}(p)$, which should vanish in order to maintain (vector) gauge-invariance.  Again we use W-T identities to simplify the expression, and we find
\beq 
k_{1\mu} T^{\lambda \mu \nu}(q,k_1,k_2;p) &=& \nonumber \\
g  \int \frac{d^4 p}{(2 \pi)^4} \ {\rm tr} \Big[\hspace{-10pt}&&\hspace{-18pt}\gamma^5 \Delta(p-q) \Gamma^\nu(p,-k_2)\Delta (p)\Gamma^\lambda (p-q,q)  \nonumber \\
&-&\hspace{-10pt} \gamma^5 \Delta(p-k_2)\Gamma^\nu(p,-k_2)\Delta (p) \Gamma^\lambda (p-q,q) \nonumber \\
&+&\hspace{-10pt}\gamma^5  \Delta(p-q)\Gamma^\nu(p-k_1,-k_2)\Delta (p-k_1) \Gamma^\lambda (p-q,q) \nonumber \\
&-&\hspace{-10pt} \gamma^5  \Delta(p-q)\Gamma^\nu(p-k_1,-k_2)\Delta (p) \Gamma^\lambda (p-q,q) \nonumber \\
&+&\hspace{-10pt} \gamma^5 \Delta(p-q)\Gamma^\nu(p-k_1,-k_2)\Delta (p) \Gamma^\lambda (p-q,q) \nonumber \\
&-&\hspace{-10pt} \gamma^5\Delta(p-q)\Gamma^\nu(p,-k_2)\Delta(p)\Gamma^\lambda(p-q,q) \nonumber \\
&+&\hspace{-10pt} \gamma^5 \Delta(p-k_2)\Gamma^\nu(p,-k_2)\Delta (p) \Gamma^\lambda (p-q,q) \nonumber \\
&-&\hspace{-10pt} \gamma^5 \Delta(p-k_2)\Gamma^\nu(p,-k_2)\Delta (p) \Gamma^\lambda (p-k_2,q) \Big].
\eeq
Here we have exact cancellation between the first and sixth terms,  the fourth and fifth, and the second and seventh.  We also see that the third term generates the eighth term by taking $p\rightarrow p+k_1$, so we have
\beq
k_{1\mu} T^{\lambda \mu \nu}(q,k_1,k_2;p)=-g  \int \frac{d^4p}{(2 \pi)^4} \big[f_V(p+k_1)-f_V(p)\big]
\eeq
where
\beq
f_V(p) &=& {\rm tr} \left[\gamma^5 \Delta(p-q) \Gamma^\nu(p-k_1,-k_2) \Delta(p-k_1)\Gamma^\lambda(p-q,q)\right] \nonumber \\
&\supset& \frac{-4ig^2 \epsilon^{\lambda \nu \alpha \beta}p_\alpha k_{2 \beta}}{p^4}.
\eeq 
Finally we find
\beq
k_{1\mu} T^{\lambda \mu \nu}(q,k_1,k_2;p)=\frac{g^3}{8 \pi^2}\epsilon^{\lambda \nu \alpha \beta}k_{1\beta}k_{2 \alpha}.
\eeq

The remaining step in the calculation is to account for the ambiguity in the momentum routing.  To this end we define
\beq
A^{\lambda \mu \nu} (q, k_1, k_2; p+a) \equiv T^{\lambda \mu \nu}(q, k_1, k_2; p+a)-T^{\lambda \mu \nu}(q, k_1, k_2; p).
\eeq
Again we write integral in the form of (\ref{integralform}), with $f(p)$ replaced by $f_A(p)$: 
\beq
f_A(p)&=&-{\rm tr} \big[\Gamma^{\lambda}(p-q,q)\gamma^5 \Delta(p-q) \Gamma^\mu(p-k_2,-k_1)\Delta(p-k_2)\Gamma^\nu(p,-k_2)\Delta(p) \nonumber \\
&+&\Gamma^{\lambda}(p-q,q)\gamma^5\Delta(p-q) \Gamma^\nu(p-k_1,-k_2)\Delta(p-k_1)\Gamma^\mu(p,-k_1)\Delta(p) \nonumber \\
&+&\Gamma^{\lambda}(p-q,q)\gamma^5\Delta(p-q)\Gamma^{\mu \nu}(p,-k_1,-k_2)\Delta(p) \nonumber \\
&+&\Gamma^{\lambda \mu}(p-k_2,q,-k_1)\gamma^5\Delta(p-k_2)\Gamma^\nu(p,-k_2)\Delta(p) \nonumber \\
&+&\Gamma^{\lambda \nu}(p-k_1,q,-k_2)\gamma^5\Delta(p-k_1)\Gamma^\mu(p,-k_1)\Delta(p) \big].
\eeq
The second term is related to the first by taking $k_1\leftrightarrow k_2$ and $\mu \leftrightarrow \nu$; the same relation holds for the fourth and fifth terms.  Keeping only  the linearly divergent  ${\mathcal O}(p^{-3})$ terms we find contributions only from the first two terms, where
\beq
{\rm tr}\big[\Gamma^{\lambda}(p-q,q)\gamma^5 \Delta(p-q) \Gamma^\mu(p-k_2,-k_1)\Delta(p-k_2)\Gamma^\nu(p,-k_2)\Delta(p)\big] \supset \frac{4g^3i \epsilon^{\lambda \mu \nu \alpha}p_\alpha}{p^4}
\eeq
leads to 
\beq
A^{\lambda \mu \nu} (q, k_1, k_2; p+a)=\frac{g^3\epsilon^{\lambda \mu \nu \rho} a_\rho}{8 \pi^2}+\{k_1\leftrightarrow k_2; \mu \leftrightarrow \nu\}.
\eeq
If we denote $a=\alpha k_1+(\alpha-\beta)k_2$ we find the following
\beq
A^{\lambda \mu \nu} (q, k_1, k_2; p+a)=\frac{g^3 \beta  \epsilon^{\lambda \mu \nu \rho}}{8 \pi^2}(k_1-k_2)_\rho.
\eeq

Just as in the canonical case we ensure conservation of the $U(1)_V$ symmetry by considering 
\beq
k_{1 \mu} T^{\lambda \mu \nu}(q, k_1, k_2; p+a) &=& k_{1\mu}A^{\lambda \mu \nu}(q, k_1, k_2; p+a) + k_{1\mu}T^{\lambda \mu \nu}(q, k_1, k_2; p) \nonumber \\
&=& \frac{g^3 \epsilon^{\lambda \nu \alpha \beta}}{8 \pi^2} k_{1 \beta}k_{2 \alpha}(1-\beta),
\eeq
thus to conserve the vector current we need to take $\beta =1$.
The anomaly factor is now  determined as usual:
\beq
q_\lambda T^{\lambda \mu \nu}(q, k_1, k_2; p+a) &=& q_\lambda A^{\lambda \mu \nu}(q, k_1, k_2; p+a) +  q_\lambda T^{\lambda \mu \nu}(q, k_1, k_2; p) \nonumber \\
&=& \frac{g^3 \epsilon^{\mu \nu \alpha \beta}}{4 \pi^2} k_{1 \beta}k_{2 \alpha}(1+\beta) \nonumber \\
&=& \frac{g^3 \epsilon^{\mu \nu \alpha \beta}}{2 \pi^2} k_{1 \beta}k_{2 \alpha}.
\eeq

\section{Conclusions}
We have shown that the scaling dimension of the fermion in the loop of a VVA diagram does not enter the anomaly factor.  While this might have seemed an intuitive result, we actually see that the calculation relies on non-trivial cancellations coming from Feynman diagrams that are inherent to unparticles, i.e. the cancellation is not present if only the usual  diagrams (with fermions replaced by unfermions) are used.

This result might have been anticipated from the fact that there are no higher order contributions to the VVA diagrams so that there is no renormalization of the anomaly factor at any order in perturbation theory.  So the proof that the anomaly doesn't depend on the scaling dimension provides a consistency check for the unparticle formalism.  Beyond this, this result can be viewed as a cross check of Seiberg duality which uses anomaly  matching between different unfermion loop diagrams.  

The analysis of unfermions in Seiberg duality leads to one interesting speculation.  Since the anomaly matching also includes quarks and dual quarks that have nontrivial scaling dimensions, it seems that the  unparticle Feynman rules we've used, and hence the anomaly calculation given here, might work for these unfermions as well.
However since quark fields  are not gauge invariant operators they can violate the unitarity bound, and indeed a naive application  of Eq. (\ref{rcharge}) gives a scaling dimension $d_q<3/2$. One might object that since quarks are not gauge invariant, their scaling dimensions will be gauge dependent.  However, in SUSY QCD, because of the non-renormalization of the superpotential mass term, the quark anomalous dimension is proportional to the mass anomalous dimension, which is itself gauge invariant.  Since the quark scaling dimension is thus gauge invariant (when using a supersymmetric regulator and supersymmetric gauge fixing), it may indeed make sense to extend the uparticle analysis to these non-gauge invariant fields. 

  A further problem with unquark propagators might seem to be that the spectral density in
Eq.~(\ref{prop ansatz}) is ill-defined for $d<3/2$, requiring some subtraction in order to be sensible.  On the other hand the action~(\ref{p action}) and its corresponding propagator do not seem to suffer a readily identifiable pathology as $d$ goes below $3/2$.  With this, there seems to be no obstacle in applying the unparticle analysis to quarks and dual quarks in SUSY QCD.

\section*{Acknowledgments}

We thank Giacomo Cacciapaglia, Zacharia Chacko, Joe Kiskis, Guido Marandella, Damien Martin, Bob McElrath, and David Stancato  for useful discussions and comments.  This work is supported by the US Department of Energy under contract DE-FG03-91ER40674.


\begin{thebibliography}{99}

\bibitem{Georgi1}
H.~Georgi, 
  Phys. Rev. Lett. {\bf 98} 221601 (2007)
  {\tt [arXiv:hep-ph/0703260]}.

\bibitem{Georgi2}
H.~Georgi,
  Phys.\ Lett.\  B {\bf 650}, 275 (2007)
  {\tt [arXiv:0704.2457 [hep-ph]]}.

\bibitem{shirmanquiros}
  P.~J.~Fox, A.~Rajaraman and Y.~Shirman,
  Phys.\ Rev.\  D {\bf 76} (2007) 075004
  {\tt [arXiv:0705.3092 [hep-ph]]};
A.~Delgado, J.~R.~Espinosa and M.~Quiros,
  JHEP {\bf 0710} (2007) 094
  {\tt [arXiv:0707.4309 [hep-ph]]}.

\bibitem{uncolor}
G.~Cacciapaglia, G.~Marandella and J.~Terning,
  JHEP {\bf 0801}, 070 (2008)
  {\tt [arXiv:0708.0005 [hep-ph]]}.
  
  

 
  


  
  

\bibitem{unfermions}
  G.~Cacciapaglia, G.~Marandella and J.~Terning,
  {\tt arXiv:0804.0424 [hep-ph]}.

  
\bibitem{ABJ}
S.~L.~Adler,
  Phys.\ Rev.\  {\bf 177}, 2426 (1969);
J.~S.~Bell and R.~Jackiw,
  Nuovo Cim.\  A {\bf 60}, 47 (1969).
  
\bibitem{Fujikawa}
  K.~Fujikawa,
  Phys.\ Rev.\ Lett.\  {\bf 42}, 1195 (1979);
  Phys.\ Rev.\  D {\bf 21}, 2848 (1980)
  [Erratum-ibid.\  D {\bf 22}, 1499 (1980)].
  



\bibitem{Seiberg}
  N.~Seiberg,
  Nucl.\ Phys.\  B {\bf 435}, 129 (1995)
  {\tt [arXiv:hep-th/9411149]}.



\bibitem{tHooft}
  G.~'t Hooft,
  NATO Adv.\ Study Inst.\ Ser.\ B Phys.\  {\bf 59}, 135 (1980).

\bibitem{AdlerBardeen}
  S.~L.~Adler and W.~A.~Bardeen,
  Phys.\ Rev.\  {\bf 182} (1969) 1517.
  
  
\bibitem{Neubert}
  M.~Neubert,
  Phys.\ Lett.\  B {\bf 660}, 592 (2008)
  {\tt [arXiv:0708.0036 [hep-ph]]}.


   \bibitem{Mack}
G.~Mack, 
{\em Comm. Math. Phys.} 55 (1977) 1.


\bibitem{reviews}
For reviews of the contraints on scaling dimensions in conformal theories see,
  S.~Minwalla,
  Adv.\ Theor.\ Math.\ Phys.\  {\bf 2}, 781 (1998)
  {\tt [arXiv:hep-th/9712074]};
       J.~Terning,
  ``Modern supersymmetry: Dynamics and duality,''
{\it  Oxford, UK: Clarendon (2006) p 128}.

    
 
\bibitem{Flato}
M.~Flato and C.~Fronsdal,
{\em Lett.\ Math.\ Phys.} 8 (1984) 159;
V.~K.~Dobrev and V.~B.~Petkova,
{\em Phys.\ Lett.\ B} 162 (1985) 127;
V.~K.~Dobrev and V.~B.~Petkova, 
{\em Symposium on
Conformal Groups and Structures, Clausthal 1985: Lecture Notes in Physics, Vol.
261,}
Eds. Barut, A.~O.~and Doebner, H.~D.~ (Springer-Verlag, Berlin, 1986) p. 300;
V.~K.~Dobrev and V.~B.~Petkova,
{\em Fortsch.\ Phys.}\ 35 (1987) 537.

  
  
\bibitem{Basu}
  R.~Basu, D.~Choudhury and H.~S.~Mani,
  {\tt arXiv:0803.4110 [hep-ph]}.
  
\bibitem{Cheung}
K.~Cheung, W.~Y.~Keung and T.~C.~Yuan,
  Phys.\ Rev.\ Lett.\  {\bf 99}, 051803 (2007)
  {\tt [arXiv:0704.2588 [hep-ph]]}.
  


\bibitem{Grinstein}
  B.~Grinstein, K.~Intriligator and I.~Z.~Rothstein,
  {\tt arXiv:0801.1140 [hep-ph]}.
 
\bibitem{minimal}
J.~Galloway, D.~Martin and D.~Stancato,
  {\tt arXiv:0802.0313 [hep-th]}.
  
\bibitem{NL}
  M.~Chretien and R.~E.~Peierls,
  Proc.\ Roy.\ Soc.\ Lond.\  A {\bf 223} (1954) 468.

\bibitem{NCQM}
J.~Terning
  Phys. Rev. D {\bf 44} (1991) 887.
 

\bibitem{WT}
  J.C.~Ward, Phys. Rev. {\bf 78} 1824 (1950); Y.~Takahashi, Nuovo Cimento {\bf 6} 370 (1957).


\bibitem{Bertlmann}
        R.~A.~Bertlmann,
        ``Anomalies in quantum field theory,''
      (Oxford, UK: Clarendon, 1996) p. 566.




\end{thebibliography}
\end{document}